\documentstyle[sprocl,epsfig]{article}

\def\be{\begin{equation}}
\def\ee{\end{equation}}
\def\bea{\begin{eqnarray}}
\def\eea{\end{eqnarray}}
\begin{document}

\title{OBSERVATION OF ATMOSPHERIC NEUTRINO EVENTS WITH THE AMANDA EXPERIMENT}

%\address{Dept. of Physics, University of Wisconsin, 1150 University Avenuw, 
%Madison, WI, 53706, USA, 
%e-mail: karle@alizarin.physics.wisc.edu}

\author{Abrecht Karle
\footnote{Talk presented at the 17th International Workshop on Weak Interactions 
 and Neutrinos (WIN99), Cape Town, South Africa, January 1999 },
for the AMANDA collaboration}

\author{
\vskip 0.4 cm 
E. Andres$^{8}$, 
P. Askebjer$^{4}$, 
G. Barouch$^{8}$, 
S.W. Barwick$^{6}$, 
X. Bai$^{8}$, 
K.  Becker$^{9}$, 
R. Bay$^{5}$, 
L. Bergstr\"om$^{4}$, 
D. Bertrand$^{12}$, 
D. Besson$^{13}$, 
A. Biron$^{2}$, 
J. Booth$^{6}$, 
O. Botner$^{14}$, 
A. Bouchta$^{2}$, 
S. Carius$^{3}$, 
M. Carlson$^{8}$, 
W. Chinowsky$^{10}$, 
D. Chirkin$^{5}$, 
J. Conrad$^{14}$, 
C. Costa$^{8}$, 
D. Cowen$^{7}$, 
E. Dalberg$^{4}$, 
J. Dewulf$^{12}$, 
T. DeYoung$^{8}$, 
J. Edsj\"o$^{4}$, 
P. Ekstr\"om$^{4}$, 
G. Frichter$^{13}$, 
A. Goobar$^{4}$, 
L. Gray$^{8}$, 
A. Hallgren$^{14}$, 
F. Halzen$^{8}$, 
Y. He$^{5}$, 
R. Hardtke$^{8}$, 
G. Hill$^{8}$, 
P.O. Hulth$^{4}$, 
S. Hundertmark$^{2}$, 
J. Jacobsen$^{10}$, 
V. Kandhadai$^{8}$, 
A. Karle$^{8}$, 
J. Kim$^{6}$, 
B. Koci$^{8}$, 
M. Kowalski$^{2}$, 
I. Kravchenko$^{13}$, 
J. Lamoureux$^{10}$, 
P. Loaiza$^{14}$, 
H. Leich$^{2}$, 
P. Lindahl$^{3}$, 
T. Liss$^{5}$, 
I. Liubarsky$^{8}$, 
M. Leuthold$^{2}$, 
D.M. Lowder$^{5}$, 
J.  Ludvig$^{10}$, 
P. Marciniewski$^{14}$, 
T. Miller$^{1}$, 
P. Miocinovic$^{5}$, 
P. Mock$^{6}$, 
M. Newcomer$^{7}$, 
R. Morse$^{8}$, 
P. Niessen$^{2}$, 
D. Nygren$^{10}$, 
C. P\'erez de los Heros$^{14}$, 
R. Porrata$^{6}$, 
P.B. Price$^{5}$, 
G. Przybylski$^{10}$, 
K. Rawlins$^{8}$, 
W. Rhode$^{5}$, 
S. Richter$^{2}$, 
J. Rodriguez Martino$^{4}$, 
P. Romenesko$^{8}$, 
D. Ross$^{6}$, 
H. Rubinstein$^{4}$, 
E. Schneider$^{6}$, 
T. Schmidt$^{2}$, 
R. Schwarz$^{8}$, 
A. Silvestri$^{2}$, 
G. Smoot$^{10}$, 
M. Solarz$^{5}$, 
G. Spiczak$^{1}$, 
C. Spiering$^{2}$, 
N. Starinski$^{8}$, 
P. Steffen$^{2}$, 
R. Stokstad$^{10}$, 
O. Streicher$^{2}$, 
I. Taboada$^{7}$, 
T. Thon$^{2}$, 
S. Tilav$^{8}$, 
M. Vander Donckt$^{12}$, 
C. Walck$^{4}$, 
C. Wiebusch$^{2}$, 
R. Wischnewski$^{2}$, 
K. Woschnagg$^{5}$, 
W. Wu$^{6}$, 
G. Yodh$^{6}$, 
S. Young$^{6}$
}

\address{
\begin{itemize}
\item[(1)] Bartol Research Institute, University of Delaware, Newark, DE, USA
\item[(2)] DESY-Institute for High Energy Physics, Zeuthen, Germany
\item[(3)] Dept. of Physics, Kalmar University, Sweden
\item[(4)] Dept. of Physics, Stockholm University, Stockholm, Sweden
\item[(5)] Dept. of Physics, UC Berkeley, Berkeley, CA, USA
\item[(6)] Dept. of Physics, UC Irvine, Irvine, CA, USA
\item[(7)] Dept. of Physics, University of Pennsylvania, Philadelphia, PA, USA
\item[(8)] Dept. of Physics, University of Wisconsin, Madison, WI, USA
\item[(9)] Dept. of Physics, University of Wuppertal, Wuppertal, Germany
\item[(10)] Lawrence Berkeley Laboratory, Berkeley, CA, USA
\item[(11)] South Pole Station, Antarctica
\item[(12)] ULB - IIHE - CP230,  Boulevard du Triomphe, B-1050 Bruxelles
\item[(13)] University of Kansas, Lawrence, KS, USA
\item[(14)] University of Uppsala, Uppsala, Sweden
\end{itemize}
}

\maketitle\abstracts{
A first analysis of the AMANDA-B 10-string array data is presented.
A total of 113 days of data from its first year of operation in 1997 have
been analyzed. High energy neutrinos generate upward moving muons.
Cosmic ray muons penetrating the ice sheet to a depth of
2000\,m are the major source of background. We discuss the
method used to reject the background of approximately $0.5\cdot 10^9$ 
downgoing muons and leave 17 upward going events.
The neutrino candidates are discussed and compared with expectations.
}

\section{Rejection of atmospheric muon background}

In the Austral summer 96-97 the construction of the 
first generation AMANDA detector was completed. The detector consists
of 300 optical sensors on 10 strings located at depths of 1500 to 2000\,m
in the deep Antarctic ice. The calibration and the performance 
characteristics of the AMANDA array are described in reference \cite{b4}.
In this report we present a first analysis of data taken during a period 
of 113 days during the first year of operation in 1997. The detector
live time corresponds to about 85 days of data.

\begin{table}[htbp]
\vskip -0 cm 
\caption{Rejection of background and efficiency for atmospheric neutrinos  
at background rejection levels 1 to 4. The meaning of the cuts is explained 
in the text. Two categories of "direct hits" are used: B) [-5,+25]\,nsec, 
and C) [-5,+75]\,nsec. The results are given for a Monte-Carlo simulation of 
cosmic ray muons (14\,h), for a simulation of atmospheric neutrinos (85\,d), 
and for experimental data (85\,d). 
}
\label{tab_all_in_one}
\begin{center}
\begin{tabular}{|l|c|c|c|c|c|c|c|}
\hline
 &  Cut Level                 &    0  &    1    &    2  &    3    &    4   \\ %[1ex]
\hline
Filter &  filter   &       &  yes    &  yes  &  yes    &  yes   \\ %[1ex]
\hline
Quality &  Direct B Hits      &       &        &$\geq$ 5&$\geq$ 5 &$\geq$ 6 \\ %[1ex]
cuts &  Direct C Hits         &       &         &       &$\geq$ 10 &$\geq$ 15 \\ %[1ex]
 &  D. Length [m]             &       &         &       &$\geq$ 100 &$\geq$ 100  \\ %[1ex]
 &  Edge cut                  &       &         &       &         &  yes    \\ %[1ex]
\hline
Zenith & $ \theta_1$ (line fit) &       &         &       & $\geq 80^\circ$  & $\geq 100^\circ$ \\ %[1ex]
angle & $\theta_2$ (full fit)  &       & $\geq80^\circ$ & $\geq 80^\circ$ 
                                            & $\geq 100^\circ$ & $\geq 100^\circ$ \\ %[1ex]
\hline
Results &  MC: atmos. $\mu$   &  $3.4\cdot 10^6$ &   $2.1\cdot 10^5$  &  853            &  0 &  0   \\ %[1ex]
 &  MC: atmos. $\nu$      &    2000          &   1016             &  272            & 89 & 21.1 \\ %[1ex]
 &  Exp. Data                      & $4.9\cdot 10^8$  & $4.5\cdot 10^7$    & $3.5\cdot 10^5$ & 452 & 17  \\ %[1ex]
\hline
\end{tabular}
\end{center}
\end{table}

Atmospheric muons are recorded at a rate of 70\,Hz. Upward going
atmospheric muon neutrinos 
are expected to trigger the AMANDA 10-string detector at a rate of about
$3\cdot 10^{-4}$\,Hz or 25 events per day. 
The only parameter for background rejection is the 
direction of the reconstructed track, which decides whether a muon was 
moving upward or downward. 
Upward muon tracks are generated by neutrinos, where downward moving 
tracks are totally dominated by penetrating cosmic ray muons 
generated in the atmosphere.
About 90\% of the cosmic ray muons are rejected with a 
simple filter method based on the correlation of arrival times and depth 
of the observed Cherenkov photons.
The remaining events are reconstructed by fitting 
the Cherenkov light cone generated by a relativistic particle 
to the observed arrival times \cite{wiebusch}. 
After the initial reconstruction a set of quality cuts are applied 
to suppress a remaining background of muons which were reconstructed 
as upward moving. 
The most important cut is the number of "direct hits" in an event.
A direct hit is a photon that is detected within a time interval
of [-5,+25] nsec of the fitted Cherenkov cone. 
Another important criterion is the "direct length" cut which 
requires that the direct hits are distributed over a muon track of at 
least 100\,m length.
A combination of two other cuts is the "edge cut" which requires 
that the event was not exclusively 
concentrated at the top or bottom edge of the detector.

\section{Observation of atmospheric neutrino candidates}

\begin{figure}
\vskip -1.7 cm 
\hbox{
\hskip -0.3cm\vbox{
\epsfxsize=2.4in
\epsffile{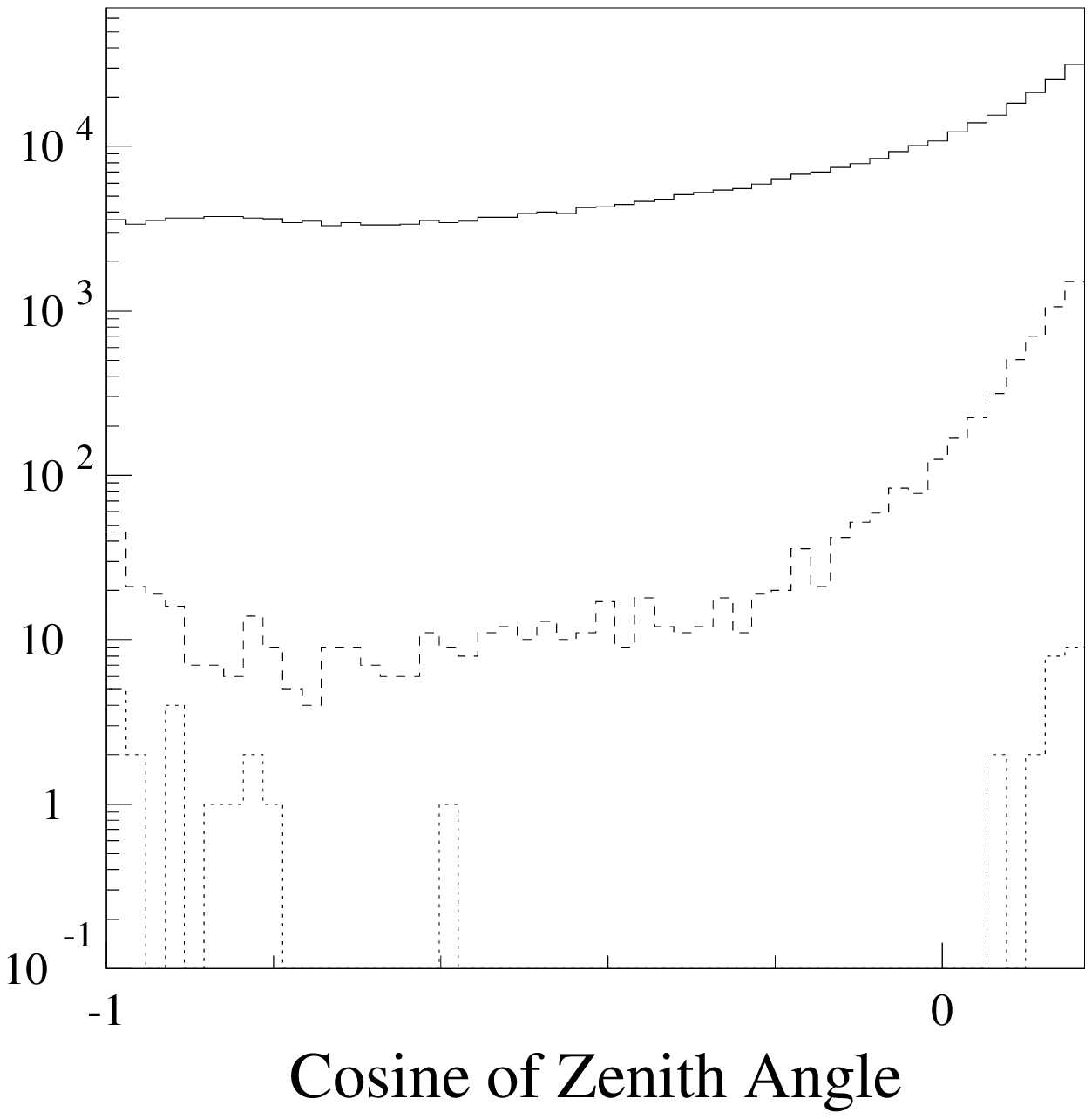}
}
\hskip -0.2 cm \raise 1.9cm \vbox{
\epsfxsize=2.2in
\epsffile{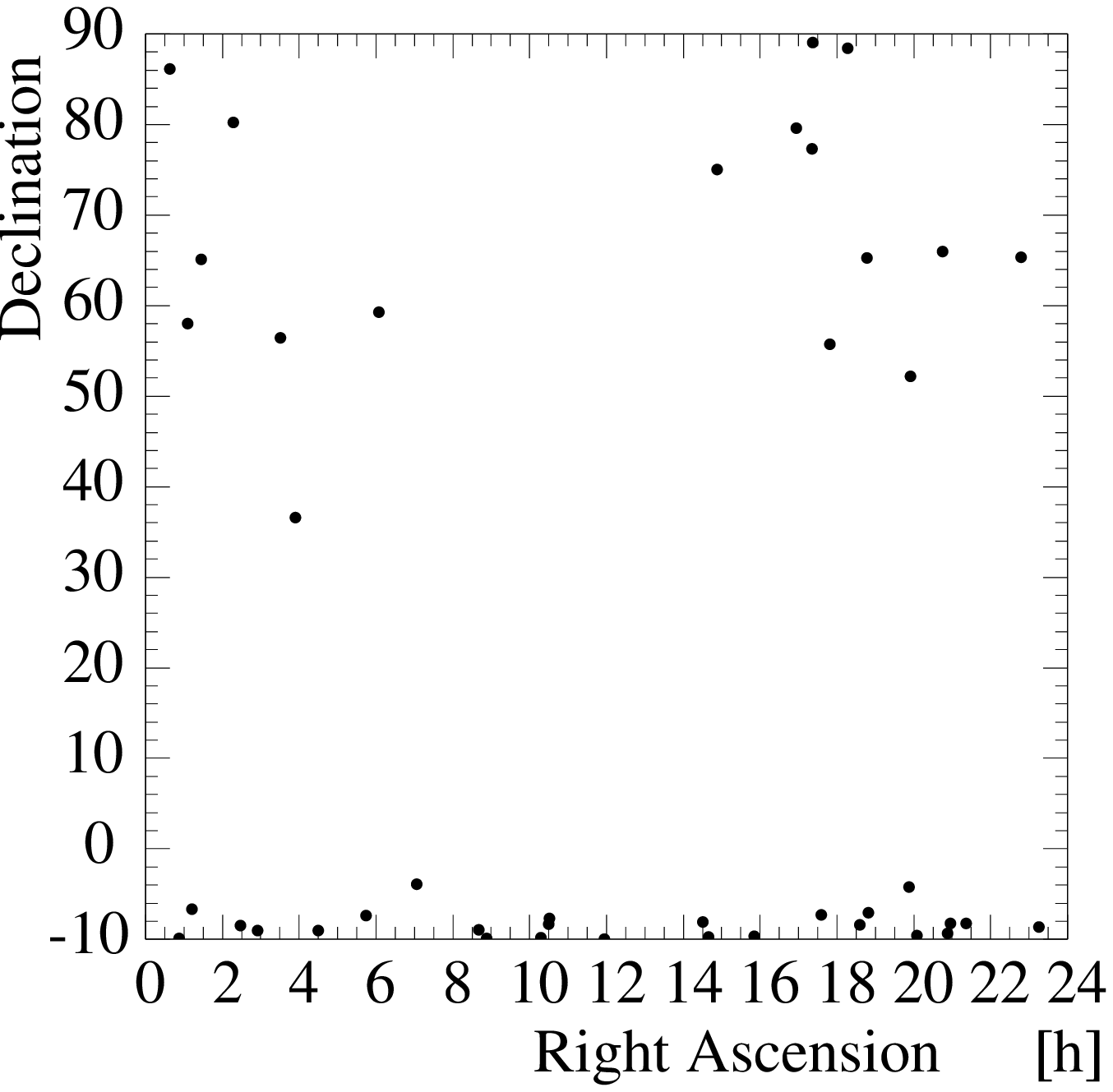}
}
}
\vskip -2.3 cm 
\caption{ The reconstructed zenith angles of 113 days of AMANDA 10 string data is shown 
for quality level 2 (solid lines), 3 (dashed) and 4 (dotted).
The plot on the right shows the sky plot of all events that pass level 4 quality cuts.}
\label{zenith_overlay_cos}
\end{figure}

\begin{figure}
\vskip -0.5 cm 
\hbox{
\hskip -0.2cm  \raise 0.3cm 
\vbox{
\epsfxsize=2.2in
\epsffile{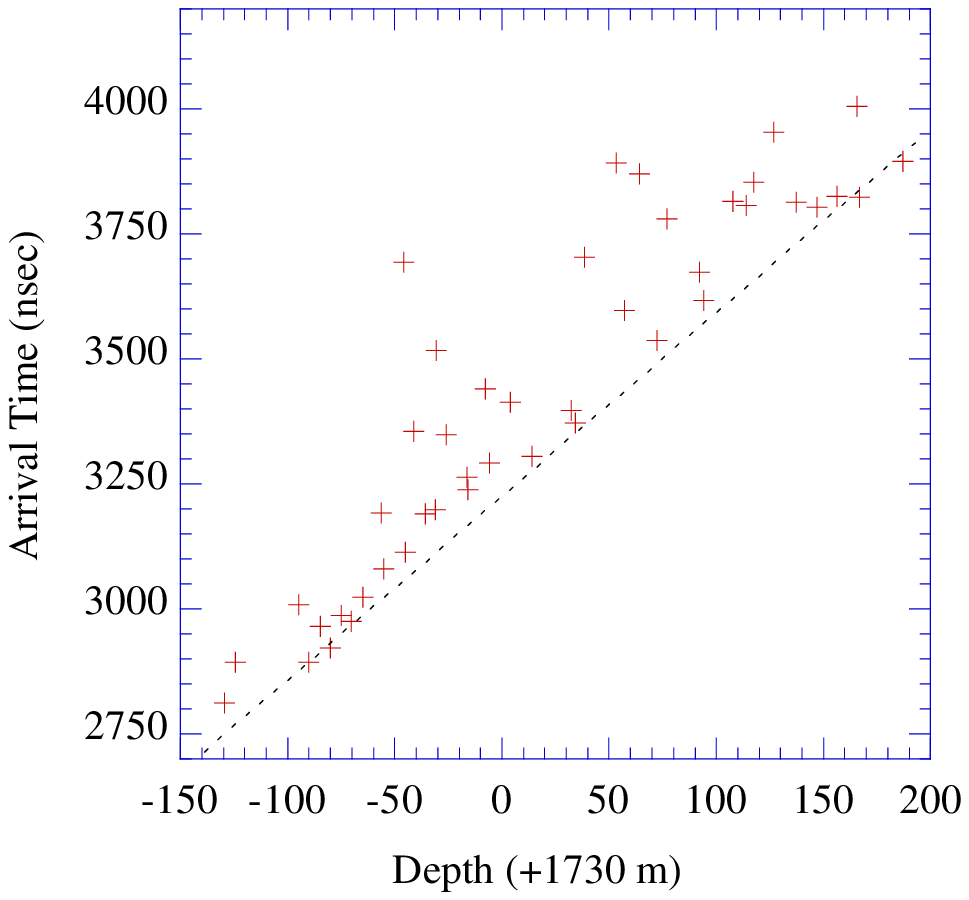}
}
\hskip -0.2 cm \raise 0.0cm \vbox{
\epsfxsize=2.4in
\epsffile{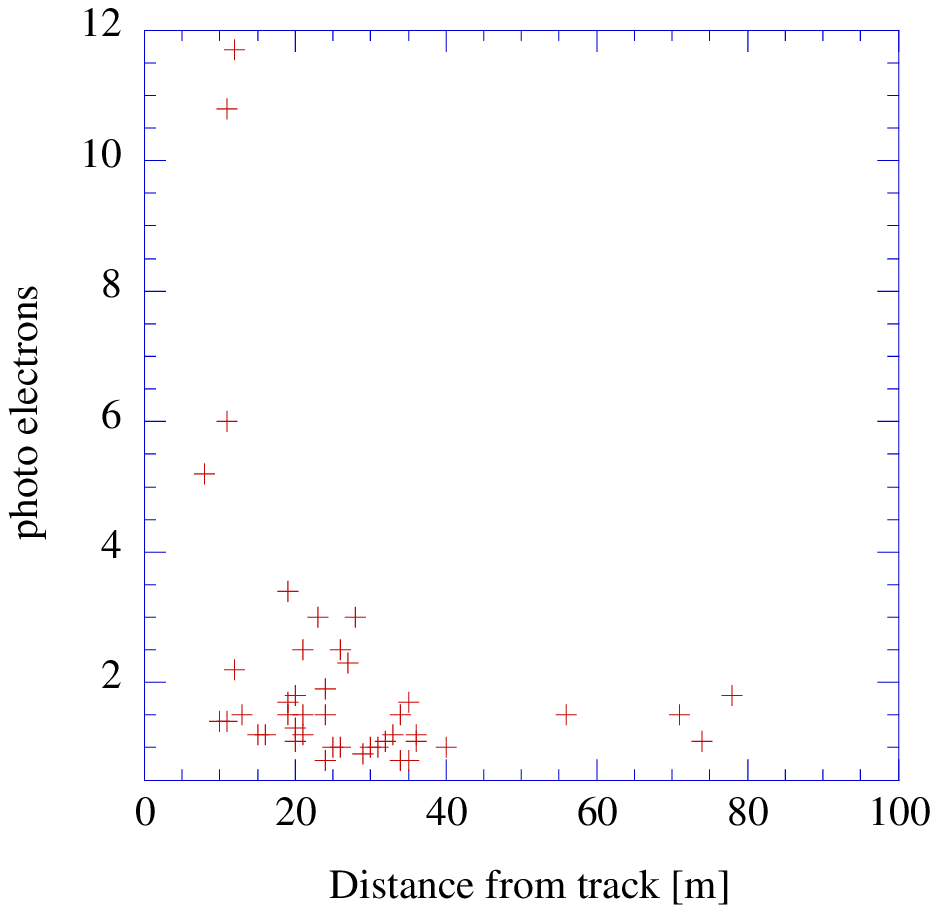}
}
}
\vskip -0.5 cm 
\caption{ Event 1197960: The recorded arrival time of photons 
is plotted versus the depth of the observing sensors (left). The slope of the 
dashed line is the result of the reconstructed zenith angle of $155^\circ$.
The observed pulse amplitudes are plotted versus 
distance of the track (right). }
\label{ArrTimeVsDepth_1197960}
\vskip -0.3 cm 
\end{figure}

We reduce the cosmic ray muon background in four steps, to which we 
refer as rejection level 1 to 4. The definitions of the four cut levels
are summarized in table \ref{tab_all_in_one}.
Figure \ref{zenith_overlay_cos} shows the distribution of 
the reconstructed zenith angles up to $10^\circ$ above the horizon 
for the applied quality cuts from level 2 to 4.
In the same figure the sky coordinates of the remaining 17 events 
are shown for quality level 4. 
Above the horizon the tail of downgoing muons is visible.
However, where at cut level 2 and 3 a background of fakes is present
below the horizon, a cluster of upgoing tracks appears which is separated
from the downgoing background. 
The 17 of $4.9\cdot 10^8$ events which pass the highest
quality cuts are concentrated at larger zenith angles.
The distribution in right ascension is statistically consistent with a 
random distribution. 
A close inspection of the spatial topology and the 
amplitudes of the 17 events shows that one of the 17 events is likely to be a  
$\nu_e$ initiated cascade or a bremstrahlung event.
 The event characteristics of the remaining 16 events are in  agreement with the 
        expectation for upgoing neutrino induced muons.
A display of a neutrino candidate which extends over a length 
of 400\,m through the entire detector is shown by Halzen \cite{halzen}.
The upward moving signature of this event is illustrated 
in figure \ref{ArrTimeVsDepth_1197960} where the photon arrival times of this event 
are plotted versus the 
depth of the sensors. The slope matches the vertical velocity of 
a track reconstructed at a zenith angle of $155^\circ$, 
which agrees with the result of the full Cherenkov cone fit.
Figure \ref{ArrTimeVsDepth_1197960} also shows the amplitudes as a function of 
distance of the reconstructed muon track.  The observed photon density 
is high for sensors close to the track.

\section{Comparison with Monte-Carlo prediction and conclusion}

\begin{figure}
\begin{center}
\vskip -1.80 cm 
  \hskip 0cm
\hbox{
  \hskip -0.5 cm
  \raise 0.0cm 
  \vbox{
    \epsfysize=3.6 in
       \epsffile{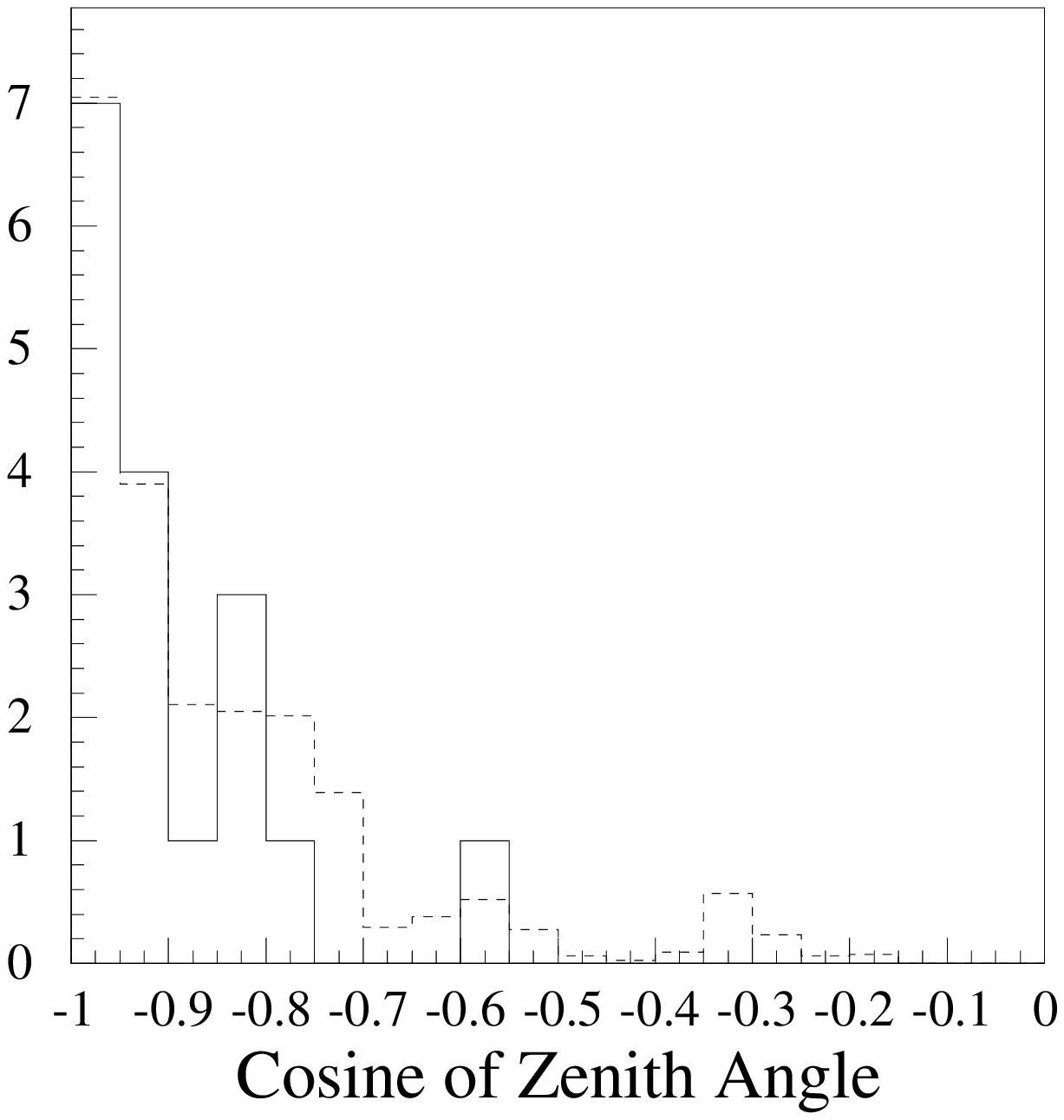}
  }
  \hskip -0.5 cm
  \raise 2.0cm 
  \vbox{
    \epsfysize=2.3in
  \epsffile{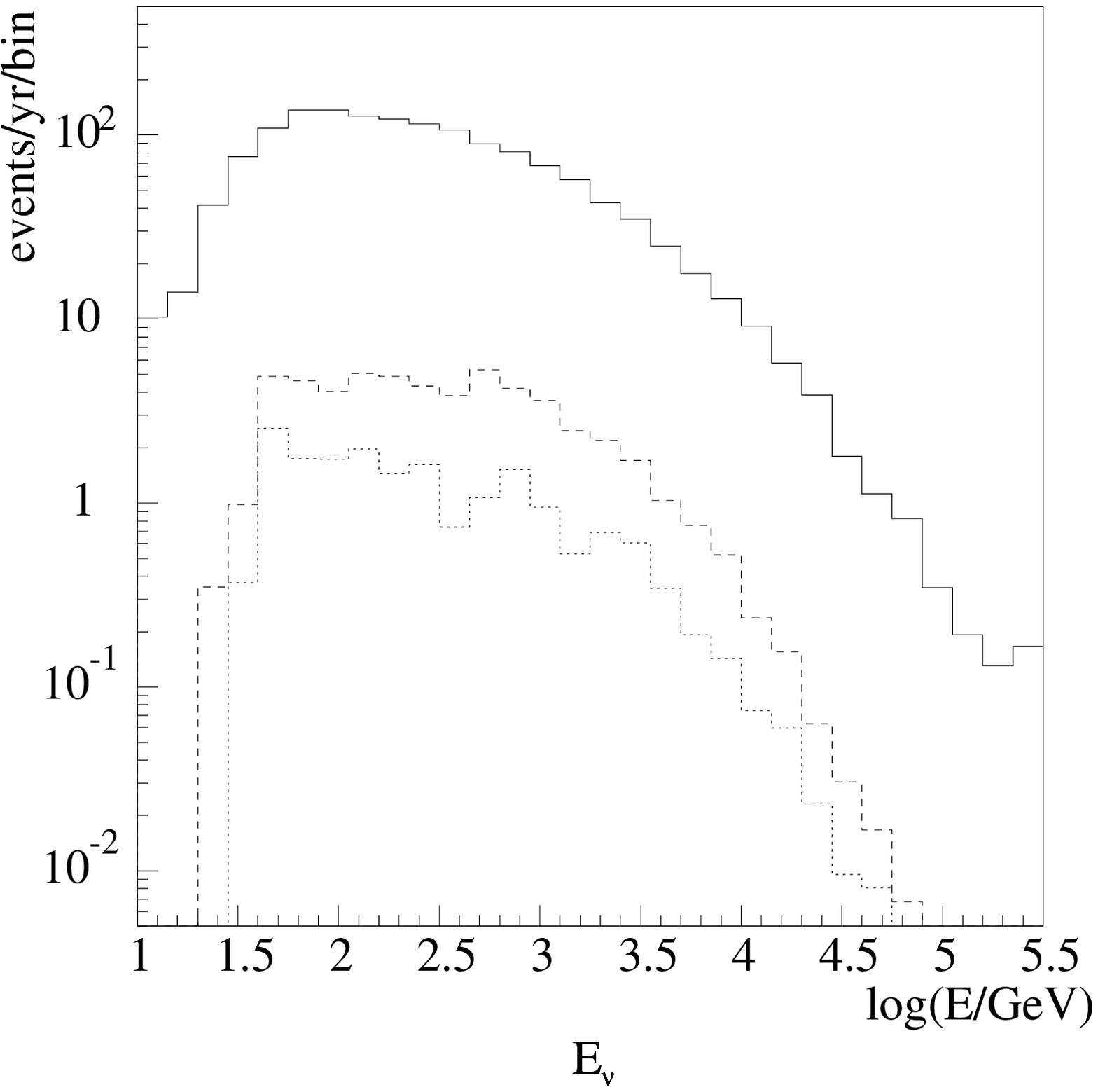}
  }
}
\vskip -2.3 cm 
\caption{Left: Zenith angle distribution of neutrino candidates and 
of MonteCarlo simulated atmospheric neutrinos.
Right: 
The simulated energy spectrum (true neutrino energy) is shown 
at trigger level of the 10 string array (solid lines),
at level 3 cuts (dashed lines) and at level 4 (dotted).}
\label{level4_zenith_overlay_cos}
\end{center}
\vskip -0.4 cm 
\end{figure}

A full simulation of atmospheric neutrinos has been performed
which predicts that 21 $\nu_\mu$ and $\overline{\nu_\mu} $ events pass
the level 4 cuts.
Figure \ref{level4_zenith_overlay_cos} shows the zenith angle
distribution of all events at level 4 along with the prediction of 
the atmospheric neutrino simulation. The energy distribution 
of simulated atmospheric neutrinos is shown for cut levels 0, 3, and 4.
The energy and angular characteristics of the atmospheric
neutrino spectrum are taken from Lipari~\cite{Lipari}. 
We estimate that the combined error of theoretical prediction and 
absolute sensitivity of the detector is 50\% or greater.
The angular distribution of the observed upward moving tracks 
 agrees well with the expectation from atmospheric neutrinos.
It illustrates the higher sensitivity of the 10 string
array to small nadir angles, reflecting that the detector is 400\,m tall,
but only 120\,m in diameter. 
Deployments in the 99-00 Antarctic summer will result in a more 
symmetric detector \cite{halzen}.

\section*{Acknowledgments}
\small

This research was supported by the following agencies:

1. U.S. National Science Foundation, Office of Polar Programs;
2. U.S. National Science Foundation, Physics Division;
3. University of Wisconsin Alumni Research Foundation; 
4. U.S. Department of Energy;
5. U.S. National Energy Research Scientific
Computing Center (supported by the Office of Energy Research
of the U.S. Department of Energy);
6. Swedish Natural Science Research Council;
7. Swedish Polar Research Secretariat;
8. Knut and Alice Wallenberg Foundation, Sweden;
9. Deutsches Elektronen-Synchrotron (DESY).

\end{document}